\documentclass[aps,pra,twocolumn,groupedaddress,showpacs,10pt]{revtex4-1}

\usepackage{graphicx}
\usepackage{epstopdf}

\begin{document}

\newcommand{\Cs}{${}^{133}\text{Cs }$}
\newcommand{\Li}{${}^{6}\text{Li }$}
\newcommand{\LiCs}{${}^{6}\text{Li} {}^{133}\text{Cs}$}
\newcommand{\tento}[1]{$\times 10^{#1}$}
\newcommand{\Csma}{$\left|3,+3\right\rangle$}
\newcommand{\Liplus}{$\left|1/2,1/2\right\rangle$}
\newcommand{\Liminus}{$\left|1/2,-1/2\right\rangle$}

\title{Observation of interspecies \Li-\Cs Feshbach resonances}

\author{M. Repp }
\author{R. Pires}
\author{J. Ulmanis}
\author{R. Heck}
\author{E. D. Kuhnle}
\author{M. Weidem\"uller}
\email[]{weidemueller@uni-heidelberg.de}
\affiliation{Physikalisches Institut, Ruprecht-Karls Universit\"{a}t Heidelberg, Im Neuenheimer Feld 226, 69120 Heidelberg, Germany}

\author{E. Tiemann}
\affiliation{Institut f\"{u}r Quantenoptik, Leibniz Universit\"{a}t Hannover, Welfengarten 1,  30167 Hannover, Germany}

\date{\today}

\begin{abstract}
We report on the observation of nineteen interspecies Feshbach resonances in an optically trapped ultracold Bose-Fermi mixture of \Cs and \Li in the two energetically lowest spin states. We assign the resonances to s- and p-wave molecular channels by a coupled-channels calculation, resulting in an accurate determination of LiCs ground state potentials. Fits of the resonance position based on the undressed Asymptotic Bound State model do not provide the same level of accuracy as the coupled-channels calculation. Several broad s-wave resonances provide prospects to create fermionic LiCs molecules with a large dipole moment via Feshbach association followed by stimulated Raman passage. Two of the s-wave resonances overlap with a zero crossing of the Cs scattering length which offers prospects for the investigation of polarons in an ultracold Li-Cs mixture.
\end{abstract}


\maketitle


The ability to control and manipulate atomic interactions via magnetically tunable Feshbach resonances (FRs) presents a unique toolbox for the field of ultracold atoms \cite{Chin2010}. FRs are used for the creation of deeply bound molecules via Feshbach association \cite{Regal2003,Herbig2003} followed by stimulated Raman adiabatic passage \cite{Ni2008,Danzl2008,Lang2008}, which gives access to the study of many-body physics, quantum chemistry and precision measurements \cite{Carr2009,Wall2012}. With a permanent electric dipole moment of 5.5 Debye  \cite{Aymar2005,Deiglmayr2010}, the largest among all alkali-metal dimers, a system of LiCs molecules in their energetically lowest states \cite{Deiglmayr2008} is  considered to be an excellent candidate for the investigation of dipolar quantum gases \cite{Pupillo2008a}. Another application of the precise tunability close to a FR is the study of Efimov trimers \cite{Braaten2007}. The large mass ratio of $m_{\text{Cs}}/m_{\text{Li}}= 22$ results in an advantageous
universal scaling factor of 4.88 instead of 22.7 as found for a system of equal masses \cite{DIncao2006}, resulting in excellent conditions for observing a series of several Efimov resonances, which, so far, has not been achieved. Interspecies FR can be also used to control the interaction between an impurity and a Bose-Einstein Condensate (BEC). Such a system can directly be mapped to the Fr\"ohlich polaron Hamiltonian \cite{Froehlich1954,Cucchietti2006,Tempere2009}, which describes the interaction of an electron gas with the charged lattice atoms in a crystal. As the excitations in a Bose-Einstein condensate represent the lattice phonons, a FR allows the precise adjustment of the modeled phonon-electron coupling strength $\alpha$, thus allowing one to explore fundamental solid-state systems.

FRs between different alkaline species have been identified for a variety of Bose-Bose \cite{Thalhammer2008,Marzok2009,Pilch2009,Cho2012},
Bose-Fermi mixtures \cite{Inouye2004,Stan2004,Ferlaino2006,Deh2008,Deh2010,Schuster2012,Park2012} as well as for the Fermi-Fermi mixture of $^6\text{Li-}^{40}\text{K}$ \cite{Wille2008}. While the broad intraspecies FRs in \Cs  \cite{Chin2004a,Berninger2011} and \Li \cite{Jochim2002,OHara2002,Bartenstein2005,Ottenstein2008,Zurn2012a} are well-explored, until now, only little is known about the interspecies interaction properties of Li and Cs. Initial theoretical investigations of the ground state potential curves by \textit{ab initio} calculations \cite{Korek2000, Aymar2005} could be improved by spectroscopic data based on Fourier-transform spectroscopy \cite{Staanum2007} and photoassociation spectroscopy \cite{Grochola2009}, mainly performed with $^{7}\text{Li} ^{133}\text{Cs}$. The elastic scattering properties of $^{7}$Li-$^{133}$Cs were studied by means of thermalization measurements at zero magnetic field \cite{Mudrich2002}, however, the tunability of the interspecies scattering properties via tuning the magnetic field has yet remained unexplored. Here, we report on the observation of nineteen interspecies loss features in different spin channels of an optically trapped \Li-\Cs mixture, by scanning a homogeneous magnetic field (Feshbach spectroscopy). The magnetic field positions and widths of the observed FRs are analyzed in a full coupled-channels calculation, allowing a consistent assignment of the resonances.

 We have realized an all-optical preparation scheme to simultaneously trap an
ultracold \Li-\Cs mixture with a magnetic field control up
to 1300 G by sequentially transferring Li and Cs atoms into an optical dipole
trap. The preparation scheme for fermionic Li follows the experimental approach
described in \cite{Ottenstein2008}.
In brief, we transfer $10^6$ Li atoms in a mixture of $\left|f=1/2, m_f=+1/2\right\rangle$ and \Liminus\ states from a magneto-optical trap (MOT) containing $10^8$ atoms to a crossed optical dipole trap (ODT) ($1/\mathrm{e}^2$ waist 60 $\mu \text{m}$, crossing angle $8.5^{\circ}$, wavelength 1070 nm), which creates an effective trap depth of $U_{\text{Li}} = 1.7$ mK at a laser power of 150 W per beam. After an evaporation phase at 760 G, where the power of the ODT is decreased in a nonlinear ramp to 0.9 W, the power of the ODT is again adiabatically increased to 1.8 W in order to minimize losses of the Li  sample during the Cs preparation.

\begin{figure*}[t]
\centering
\includegraphics[width=2.0\columnwidth]{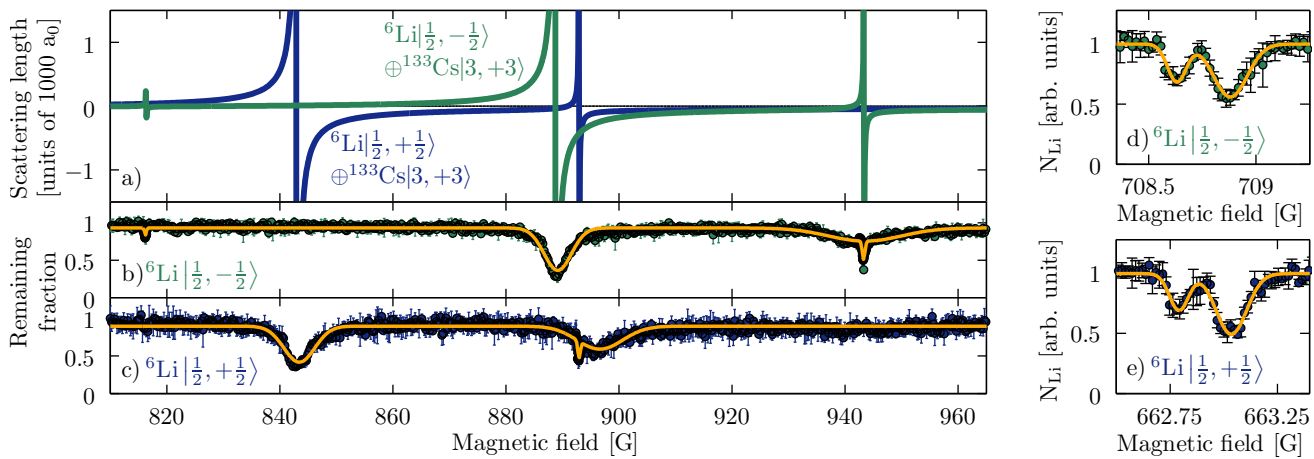}
    \caption{(Color online) Interspecies \Li-\Cs Feshbach resonances. (a) Calculated magnetic
field dependence of scattering lengths for the entrance channels
     $\text{Li}\left|1/2,1/2\right\rangle\oplus  \text{Cs}\left|3,3\right\rangle\ $(blue line)
    and $\text{Li}\left|1/2,-1/2\right\rangle\oplus
    \text{Cs}\left|3,3\right\rangle\ $(green line).
    (b) and (c) Remaining fraction of Li atoms, normalized to a sample without Cs, in the state  \Liminus \ and \Liplus , respectively, after a hold time of 100  ms at the given magnetic field. The broad loss features at 943 G in b) and 893 G  in c) arise from a Feshbach resonances  of a small amount of Cs in the $ \left|3,2\right\rangle\ $ state.
    (d) and (e) Narrow interspecies loss features with doublet structures in the
Li \Liminus \ and  Li \Liplus \ state, respectively, measured under different
conditions (see text) and normalized to the offset lines. The
data points are an average over five randomized measurements. Center positions
and widths of the loss features are obtained by fitting a sum of Gaussians (orange lines).}%
   \label{fig:Lossfeature}
\end{figure*}

After completion of the Li loading cycle, a Cs MOT is loaded slightly
displaced to the center of the ODT, in order to avoid interspecies two-body
losses \cite{Schloder1999}. During a compression phase with increased magnetic
field gradient and laser detunings as well as reduced laser powers, the sample
is superimposed with the ODT and further cooled and spin polarized by
three-dimensional degenerate Raman-sideband-cooling (DRSC)
\cite{Treutlein2001}. Reducing the ODT power to 0.9 W  during the
DRSC phase transfers the Cs atoms into the dipole trap, while the application
of an offset field after the DRSC maintains the spin polarization. After 125 ms
in the ODT, oscillations of the atomic cloud are damped out and the Cs gas is
thermalized. Since the extension of the atomic cloud during the DRSC is significantly larger than the trapping volume of the ODT, we observe the Cs temperature increase from $T_{\text{Cs}}$ =  1 $\mu$K after DRSC to $T_{\text{Cs}}$ =  8 $\mu$K during the transfer due to the non optimal spatial mode matching.

The different polarizabilities of Cs and Li for light at 1070 nm results in effective trap depths of $U_{\text{Cs}}/\text{k}_{\text{B}}= 40 \ \mu\text{K}$ and $U_{\text{Li}}/\text{k}_{\text{B}} =10 \ \mu \text{K}$  at a power of 0.9 W. We observe 5 \tento{4} Cs ({2 \tento{5}} Li) atoms at temperatures $T_{\text{Cs}}= 8 \ \mu \text{K}$ ( $T_{\text{Li}}= 2 \ \mu \text{K}$).
The trapping frequencies were measured via trap oscillations to be ${ \omega}_{\text{Cs}}/2 \pi$ = (380, 380, 30) Hz and ${ \omega}_{\text{Li}}/2 \pi$ = (900, 900, 65) Hz.
From these measurements we deduce atomic peak densities of $n_{\text{Cs}}= 3
\times 10^{11} \ \text{cm}^{-3}$  ($n_{\text{Li}}= 1 \times 10^{12} \
\text{cm}^{-3}$).
The Li sample is spin-polarized by removing one of the two $m_f$ components at high magnetic fields with a resonant light pulse. The Cs atoms are to 85\% spin polarized in the \Csma \, state  due to the DRSC, with the remaining fraction populating the $\left|3,+2\right\rangle$\ state.

Feshbach spectroscopy is performed by ramping the magnetic field to a desired value and waiting for a variable hold time between 60 and 2000 ms, which is experimentally optimized to achieve Li loss signals between 15\% and 55\%. For four very weak loss features, the Cs density  in the ODT  was furthermore increased by factors of $1.5 - 3$ by using longer loading times of the Cs MOT.
The remaining fraction of Li atoms is monitored via high-field absorption imaging providing selectivity on the spin state.
A FR manifests itself as an enhancement of atomic losses due to an increase of
the inelastic three-body recombination rate \cite{Chin2010}. In addition, the
increase in the elastic scattering cross section on a resonance leads to a
significant heating of Li by the much hotter Cs atoms, resulting in loss of
Li atoms because of the lower trap depth for this species.
Working with comparable Li and Cs atom densities, the larger visibility of FRs is thus expected by monitoring the Li atom number during the measurements.
To calibrate our magnetic field we drive a microwave (mw) transition
of Li at the magnetic fields of the maximum of the interspecies losses and determine the magnetic field from the measured mw 
frequencies with the Breit-Rabi formula. Fitting Gaussian line profiles to the mw spectra yields FWHM of 200 kHz, indicating a magnetic field accuracy of 140 mG, probably limited by the residual inhomogeneity of the magnetic field over the atomic cloud.

We find a total of nineteen loss maxima for the two  Li $\text{m}_{\text{f}}$ entrance channels, all within the range between 650 G and 950 G.
Figure \ref{fig:Lossfeature} shows four Li loss spectra in the field region,
where the broadest loss features had been obtained, as well as scans over
two narrow features, showing their doublet structure.
The loss features disappear when the Cs is removed with a resonant light pulse during the DRSC or when the measurement is performed under identical conditions with the other Li $\text{m}_{\text{f}}$ component.
Furthermore, we repeat the measurement with only Cs in the trap, to ensure that the observed Li losses are not associated with Cs intraspecies FRs, which might indirectly influence the number of optically trapped Li atoms.
Positions and widths of the losses were determined by Gaussian fits, and the results are listed in Table ~\ref{tab: List}. Trapping both spin components of Li simultaneously
with Cs atoms for 500 ms, the full accessible field range between 0 G and 1300 G was scanned with a step size of 150 mG, yielding no observable additional
interspecies loss features.

{%

\begin{table*}[t]
\newcommand{\mc}[3]{\multicolumn{#1}{#2}{#3}}

\begin{ruledtabular}
\begin{center}
\begin{tabular}{c|cc|ccc|cccc}
Entrance channel  & $ \text{B}_{\text{res}}^{\text{exp}} $[G]  & $\Delta \text{B}^{\text{exp}}$ [G]& $\delta$ [G]& $ \Delta$ [G]& $\text{a}_{\text{bg}} \ [a_{0}]$  & $\text{m}_{\text{F}}$& $\text{f}$ & G  & l \\
\hline
Li $\left|1/2,+1/2\right\rangle \ $   & 662.79(1)& 0.10(2)&-0.04  & - & - &7/2 & 9/2 &7/2 & 1 \\
 $ \oplus \ \text{Cs} \left|3,+3\right\rangle$  & 663.04(1)& 0.17(2) & -0.02 & - & - & 9/2 &9/2 & 7/2  & 1 \\
 & 713.63(2) & 0.10(3) & -0.05  & - & - &9/2& 7/2 & 7/2    & 1 \\
 & 714.07(1) & 0.14(3) & -0.05  & - & - &9/2& 7/2 & 7/2    & 1 \\
  & 843.5(4) & 6.4(1)& 0.51& 60.4 &-28.5  &7/2 & 9/2 & 7/2    & 0 \\
 & 892.87(7)* & 0.4(2) &-0.11  & 4.6 &  -28.5&7/2 & 7/2 &7/2    & 0 \\
\hline
Li $\left|1/2,-1/2\right\rangle $   & 658.21(5)& 0.2(1) &  0.07  & - & - & 7/2 & 9/2 & 7/2   & 1 \\
$ \oplus \ $Cs $\left|3,+3\right\rangle$ &708.63(1) & 0.10(2)& -0.05  & - & - & 5/2 & 7/2 & 7/2    & 1\\
 & 708.88(1) & 0.18(2)  &-0.03  & - & - &7/2 & 7/2 &7/2    & 1 \\
 &764.23(1) & 0.07(3)  & -0.06 & - & - & 7/2 & 5/2 & 7/2   & 1 \\
 &764.67(1) & 0.11(3)  &-0.05 & - & - & 7/2 & 5/2 & 7/2   & 1 \\
& 816.24(2) & 0.20(4) & -0.12 &0.36  &-28.4  & 5/2 & 9/2 &7/2   &  0 \\
 & 889.2(2) & 5.7(5)  &0.46  & 59.9 &-28.4  & 5/2 & 7/2 &7/2   & 0 \\
 & 943.26(3) & 0.38(7)  &-0.12  &4.3  & -28.4 &5/2 & 5/2 & 7/2   & 0\\
\hline
Li $\left|1/2,+1/2\right\rangle \ $  & 704.49(3) & 0.35(9) & 0.07  & - & - &7/2 & 9/2 & 7/2   & 1 \\
$\oplus \ $Cs $\left|3,+2\right\rangle$ & 896.6(7) & 10(2) & 0.68 & - &- & 5/2& 9/2& 7/2   & 0 \\
\hline
Li $\left|1/2,-1/2\right\rangle \ $  & 750.06(6) & 0.4(2)  &0.06  &  - & - & 5/2 & 7/2 & 7/2   & 1\\
$ \oplus \ $Cs $\left|3,+2\right\rangle$ & 853.85(1) & 0.15(3) & -0.17 & - &- & 3/2& 9/2 & 7/2   & 0 \\
 & 943.5(1.1)** & 15**(3)& 2.21 & - &- & 3/2 &7/2 & 7/2   & 0 \\
\end{tabular}
 \caption{List of interspecies \Li-\Cs Feshbach resonances for combinations of the entrance channels Cs $\left|f=3,m_f=+2\right\rangle$, Cs $\left|3,+3\right\rangle$, Li $\left|1/2,-1/2\right\rangle$ and Li $\left|1/2,+1/2\right\rangle$.
 The experimentally obtained positions and widths are deduced by fitting Gaussian functions with the center at the magnetic field $B_{\text{res}}^{\text{exp}}$ and the FWHM  $\Delta B^{\text{exp}}$
 to the recorded loss spectra. We deduce statistical errors by the 95\%
confidence interval of the fit. A systematic error for the center
position of $\pm$0.1 G originates from the calibration plus an additional $\pm$0.1
G from day to day drifts. The cesium density
and the hold time were varied in
order to optimize the lithium loss signal. The
resonance positions $B^{\text{theo}}_{\text{res}}$ derived from coupled channels calculations are given as deviations  $\delta$=$B_{\text{res}}^{\text{exp}}  -  B^{\text{theo}}_{\text{res}}$
with respect to the observations, showing excellent agreement. The
assigned quantum numbers  of the involved molecular states are explained in the
text. For the purely elastic channels of two body collisions at the chosen kinetic energy, the background scattering lengths
$\text{a}_{\text{bg}}$ and zero crossings of the scattering lengths  with
respect to the divergence position for 1\,nK kinetic energy is shown.
 (*) We note that the loss feature at 892.87 G overlaps with a Cs 3-body loss minimum at 893(1) G \cite{Berninger2011}. This might have a minor effect on the amount of losses due to  slightly different \Cs densities. (**) The fit is not unique, depending on the amount of data points which were considered around the small loss feature at 943.26 G. For the cc-calculation the averaged  value 942.6(1.0) G from several profile simulations was used.}
\label{tab: List}
\end{center}
\end{ruledtabular}

\end{table*}
}%


We analyze the obtained resonances in a coupled channels (cc) calculation in a similar way to previous calculations on other alkaline systems (see, e.g., \cite{Marzok2009,Schuster2012}), using the atomic parameters for Li and Cs as compiled by Arimondo \textit{et al.} \cite{Arimondo}.
Precise determination of FRs and scattering lengths relies on accurate potential curves for the $a ^3\Sigma^+$ and $X^1\Sigma^+$ states of LiCs.
Therefore, both potentials are constructed in a power series of the internuclear separation $R$, similar to the parametrization of NaK ground state curves \cite{Gerdes2008}. In a series of iterations between cc calculations for the FRs and single channel calculations of rovibrational levels within the singlet and triplet potential, the description of the potentials is improved by fitting the
underlying coefficients \footnote{The potential curves in a parametrization similar to previous work on NaK \cite{Gerdes2008} are available on request from E. Tiemann (tiemann@iqo.uni-hannover.de)}  to match the locations of the FRs and simultaneously the 6498 rovibrational transitions from laser-induced fluorescence Fourier-transform spectroscopy of the previous work \cite{Staanum2007}.

The resulting resonances are listed in Table ~\ref{tab: List} by their deviation   $\delta$=$B_{\text{res}}^{\text{exp}}  -  B^{\text{theo}}_{\text{res}}$, and the resulting scattering length dependence on the magnetic field is depicted in Figure \ref{fig:Lossfeature}.  $B^{\text{theo}}_{\text{res}}$  is the peak position of the two-body collision rate at a kinetic energy of 2 $\mu$K. The fit uses the experimental uncertainty for the weighting. The fit results in a reduced chi-squared of 0.75 for all Feshbach resonances, which is very satisfactory, since only 3 resonances are slightly beyond the experimental uncertainties. We checked that the s-wave resonances will not shift by more than $-0.02$ G, choosing a kinetic energy of the entrance channel of 20 nK instead of 2 $\mu$K, as it is adequate from the experimental conditions. The redefined potential curves reflect scattering lengths for  \Li - \Cs ($^7\mathrm{Li}$ - \Cs) of $\mathrm{a}_{\text{singlet}}=30.252(100)\ \mathrm{a}_{0}$ ($\mathrm{a}_{\text{singlet}}=45.477(150)\ \mathrm{a}_{0}$) in the singlet potential and $\mathrm{a}_{\text{triplet}}=-34.259(200)\ \mathrm{a}_{0}$ ($\mathrm{a}_{\text{triplet}}=908.6(100)\ \mathrm{a}_{0}$) in the triplet potential.

 A splitting of four p-wave resonances could be resolved in the experiment, which originates from different projection of the rotational angular momentum \textit{l} onto the space fixed axis. This splitting is described by the effective spin-spin operator $V_\mathrm{dip}( R)$ for which we used the effective form \cite{Strauss}: $V_\mathrm{dip}(R)=\frac{2}{3} \lambda(R)(3S_Z^2-S^2)$, where $S_Z$ is the total
electron spin projected onto the molecular axis.
The function $\lambda$  contains the direct magnetic spin-spin interaction as the first term and second order spin-orbit contribution as the second term
$\lambda = -\frac{3}{4}
\alpha^2\left(\frac{1}{R^3}+a_\mathrm{SO}
\exp{\left(-bR\right)}\right)$,
and is given in atomic units with $\alpha$ the universal fine structure constant. The fit resulted in $a_\mathrm{SO} = -8.0$ au and $b = 0.8$ au.

All FRs originate from the least bound levels below the atom pair asymptote 2s $^2S_{1/2}( f_{\text{Li}}=1/2)$ + 6s $^2S_{1/2}(f_{\text{Cs}}=3)$ and are significant mixtures of the triplet-singlet manifold shown by an expectation value 0.7 of the total electron spin.
Thus only the projection $\text{m}_{\text{F}}$ of the total angular momentum and
the rotational angular momentum \textit{l} are good quantum numbers.
The total angular momentum $f$, excluding \textit{l} at zero magnetic field,
and, due to the strong Cs hyperfine coupling, the angular momentum $G=S+i_{Cs}$
are also good quantum numbers which further justifies the assignment we give in
Table~\ref{tab: List}.

To characterize the resonance profiles we give the resonance
parameters for truly elastic scattering channels of two body collisions at the chosen kinetic energy:   background scattering length
$\text{a}_{\text{bg}}$ and the field difference between the resonance position
and its zero crossing  $\Delta$. Both were determined by fitting the conventional function of the scattering length as function of the magnetic field $B$:

\[
\label{resonance}
a=a_{\text{bg}}(1+\frac{\Delta}{B-B_{\text{res}}}+...),
\]
using as many terms as resonances  are present in this channel.
The channels   $\text{Li}\left|1/2,1/2\right\rangle\oplus  \text{Cs}\left|3,2\right\rangle$
and $\text{Li}\left|1/2,-1/2\right\rangle\oplus
\text{Cs}\left|3,2\right\rangle$ have significant inelastic contributions,
thus such simple representation of the scattering length as function of B will
not be appropriate.

Besides two weak and broad features at 937 G and 988 G in the s-wave entrance
channels $\text{Li}  \left|1/2,\pm1/2\right\rangle\oplus  \text{Cs}
\left|3,2\right\rangle\ $ and a very sharp feature at 1019 G for Li \Liplus \
state, the cc model suggests a series of d-wave resonances in the field region
below 400 G, which all were not detected in the scan over the full field range.
No further s- wave resonances  for magnetic fields up to 1300 G are expected for
the studied entrance channels.

We note that an assignment of the resonances via a weighted least-square fit to a simple, undressed Asymptotic Bound State Model (ABM) \cite{Wille2008,Tiecke2010} does not provide satisfactory agreement with the full set of observed resonance positions, reproducing magnetic field values of the resonances only within a range of 13 G. The ABM yields energies of the $l=0$ least bound state of $\epsilon^{0}_{0}=3323  \  \text{MHz}$ for the singlet and $\epsilon^{0}_{1}=3945 \ \text{MHz}$ for the triplet state, and for $l=1$ $\epsilon^{1}_{0}=2795 \  \text{MHz}$ for the singlet and $\epsilon^{1}_{1}=3455  \ \text{MHz}$ for the triplet state, and a singlet-triplet wavefunction overlap of $\zeta=0.759$. These values deviate significantly from the results of the cc calculation giving $\epsilon^{0}_{0}=1566 \ \text{MHz}$, $\epsilon^{0}_{1}=3942 \ \text{MHz}$, $\epsilon^{1}_{0}=1159 \ \text{MHz}$, $\epsilon^{1}_{1}=3372 \ \text{MHz}$ and $\zeta=0.866$. In order to minimize the influence of the coupling between resonant channels and continuum of the (broad) s-wave resonances, we fit only p-wave resonances, resulting in reduced deviations with $\epsilon^{0}_{0}=2546  \  \text{MHz}$, $\epsilon^{0}_{1}=3978 \ \text{MHz}$, $\epsilon^{1}_{0}=1431 \  \text{MHz}$,  $\epsilon^{1}_{1}=3626  \ \text{MHz}$ and $\zeta=0.845$. In addition, when using the values of the cc calculation in the ABM, we obtain the observed resonance spectrum, but globally shifted by about -60 G. These large discrepancies between the cc calculation, precisely reproducing all observed resonances, and the ABM indicate that an adaption of the ABM is required, which is beyond the scope of this paper.



Three of the observed s-wave resonances provide a unique tunability of the \Li-\Cs mixture.
The two resonances at 896.6 G and 889.2 G are very close to a zero crossing of
the Cs scattering length, which originates from a very broad s-wave resonance
at 787 G \cite{Lee2007,Berninger2011,Ferlaino2011}.
As the fermionic nature of \Li suppresses intraspecies collisions in a spin-polarized sample, an intriguing system featuring only tunable interspecies interactions can be created.
Such a system represents an ideal candidate for the study of the Fr\"ohlich polaron Hamiltonian \cite{Froehlich1954,Cucchietti2006,Tempere2009} with a \Li impurity in a Cs BEC. The coupling constant $\alpha $ scales with the ratio of the impurity-BEC (intra-BEC) $\text{a}_{\text{IB}}$ ( $\text{a}_{\text{BB}}$ ) scattering lengths as $\alpha\sim \text{a}_{\text{IB}}^2/\text{a}_{\text{BB}}^{4/5} $ \cite{Tempere2009}.
Self-localized phases of the polarons are predicted when the coupling extends a
critical value \cite{Cucchietti2006,Tempere2009}, which might lead to a collapse
of the BEC, if the coupling is further increased \cite{Kalas2006}. The FR at
843.5 G overlaps with the broad intraspecies FR in \Li at 832.18 G
\cite{Zurn2012a}, thus resulting in a system with
large scattering
length both between Cs and the energetically lowest Li state, as well as
between the two lowest Li states. This scenario might be used for the study of
self-localization of bosonic impurities immersed in a two-component superfluid
Fermi gas  \cite{Targonska2010}.
Additionally, the coincidence between an Efimov three-body loss minimum at 893 G \cite{Berninger2011} and two interspecies FRs allow for the realization of cooling schemes for \Li-\Cs mixtures where Cs atoms are evaporated with minimized losses \cite{Berninger2011}, while effectively cooling a Li sample sympathetically \cite{Mudrich2002}. For reaching a quantum degenerate  \Li- \Cs mixture following this approach, losses during thermalization have to be avoided. It might thus be appropriate to implement species selective optical potentials \cite{LeBlanc2007} using two different wavelengths, which would allow for creating similar trap depths for Li and Cs.

This work is supported in part by the Heidelberg Center for Quantum Dynamics. R.P. acknowledges support by the IMPRS-QD and J.U. by the DAAD.
We thank K. Meyer, S. Schmidt and R. M\"uller for their contributions in building up the experiment. We thank S. Knoop and T. Schuster for sharing parts of their ABM code and for fruitful discussions and J.P. Ang for initial calculations on the ABM model. Insightful discussions with S. Jochim and his group members are gratefully acknowledged. We thank F. Ferlaino and R. Grimm for stimulating discussions and for providing their most recent Cs Feshbach spectroscopy data to us.

\bibliography{Bib_LiCsFeshbachResonances}

\end{document}